\documentclass[a4paper,12pt]{article}
\usepackage{graphicx} 
\usepackage{comment}
\usepackage{setspace} 

\date{\empty}

\title{
\begin{flushleft}
{\large \bf
Depth resolution of transmission ERDA for H in Al under nuclear-elastically
enhanced recoiling of H by 8 MeV He  
}
\end{flushleft}
}
\begin{document}
\maketitle
\vspace{-15mm}
\author{\noindent 
H. Kudo$^{1}$
\footnote{Corresponding author\, {\it E-mail address}: kudo@tac.tsukuba.ac.jp} 
, H. Naramoto$^{1}$, M. Sataka$^{1}$, S. Ishii$^{1}$, K. Sasa$^{1, 2}$, S. Tomita$^{2}$
}
\\\\
$^{1}${\it Research Facility Center for Science and Technology, University of Tsukuba, 1-1-1 Tennodai, Tsukuba, Ibaraki 305-8577, Japan}\\
$^{2}${\it Faculty of Pure and Applied Sciences, University of Tsukuba, 1-1-1 Tennodai, Tsukuba, Ibaraki 305-8571, Japan}\\\\
ABSTRACT\\
\setstretch{1.5}

\indent Using stacked samples of Al foil and H-containing resin film, 
we have carried out elastic recoil detection analysis with transmission layout (T-ERDA)  
to investigate the depth resolution in the measurements of H distribution in Al. 
For narrow and wide acceptance conditions of the detector, the depth resolutions of 
1.5--4.9 $\mu$m at several depths in Al of 50 and 80 $\mu$m thicknesses have been 
determined for incidence of 8 MeV $^{4}$He. While the main factor to degrade the depth 
resolution is the energy straggling of recoil H for narrow acceptance conditions, 
it is the smearing of the low-energy side of H spectrum due to the angular 
spread of H for wide acceptance conditions. The knowledge obtained in this work 
is useful for analysis of 3D images of H distribution measured by T-ERDA, for example, 
future analysis of minerals or natural glass samples to determine abundances and 
distributions of water or OH in the samples.
\newpage

\section{\large Introduction}\label{intro}

\hspace{5mm} We have recently developed a non-destructive observation technique 
of hydrogen distribution in matter by applying elastic recoil detection analysis with 
transmission layout (T-ERDA) \cite{Yam2019b}.  Actually, 3D distribution of hydrogen 
bubbles of ten-$\mu$m size in Al was visualized by T-ERDA using a micro-beam.   
In this case, an 8\,MeV $^{4}$He$^{2+}$ beam collimated to a $\sim$3-$\mu$m size 
was used to scan the surface of H-charged Al. The recoil H resulting from 
hard collisions with He was energy-analyzed in the direction close to 0$^{\circ}$ 
after passing through the sample, thereby depth distribution of hydrogen was  
determined by energy-to-depth conversion using the known stopping powers. 

Under such experimental conditions, the recoil cross section as large as $\sim$2\,b/sr 
due to the nuclear elastic interaction of  8\,MeV $^{4}$He with H leads to sufficient 
count rate of recoil H \cite{Yam2019a} \cite{iba}, which allows 3D mapping of H 
using a collimated weak beam of the $\mu$m size. A remarkable advantage of T-ERDA is 
the direct detection of hydrogen in a material with a higher spatial resolution than 
for the typical case of neutron tomography \cite{Gri2014}. 
This is in contrast to other observation techniques such as electron microscopy 
(TEM, SEM) or X-ray tomography, from which information obtained is hydrogen-induced 
structural changes of the base material since these techniques are insensitive to 
hydrogen itself. 

From a different standpoint, geological samples typically contain H in the form of 
water\,(H$_{2}$O) or hydroxyls\,(OH), hence analysis of H is of scientific as well as 
technical interest. Indeed, several workers used ERDA or related techniques to 
detect hydrogen in minerals \cite{Swe1997} 
\cite{Fur2003} \cite{Kom2006} \cite{Rae2008} \cite{Bur2009} \cite{Wei2018}. 
When T-ERDA is used to observe H in precipitates or inclusions contained in the sample, 
the local stopping power associated with the local atomic composition must be used 
in the energy-to-depth conversion.  This means that we can expect to determine 
the local atomic composition, i.e., the atom species existing around H. 
The 3D images of H-containing aggregates with more than ten $\mu$m size in minerals 
or natural glass samples possibly allow to determine abundances and distributions 
of water or OH in the samples. Furthermore, such analysis   
is of geological importance since it is certainly helpful to discuss the origin 
of the lunar water and the earth's sea via hydrogen analyses of glasses and minerals 
in lunar rocks \cite{Ste2019} \cite{Lin2021}. 

With the scientific and technical background mentioned above, refinement of the  
T-ERDA technique is required to obtain detailed knowledge about 
the state or behavior of hydrogen in matter. For this purpose, we have experimentally 
determined depth resolutions for analysis of H distribution in Al.  
These data are also useful not only when Al is employed as a stopping foil for T-ERDA 
of thin samples, but also when we need to estimate spatial resolutions for H in 
other materials including mineral samples by scaling with stopping powers.  

\section{ \large Experiments }\label{exp}

\hspace{5mm} In the present experiments, Al foils and PPS [polyphenylene sulfide 
(C$_{6}$H$_{4}$S)$_{\rm n}$] films were stacked and used for test samples. 
The thicknesses of Al and PPS were determined by measuring their weight and area, 
assuming the densities of 2.70 and 1.35 g$/$cm$^{3}$ for Al and PPS, respectively. 
The measurement reproduced the nominal Al thicknesses of 5--80\,$\mu$m within 
$\pm$2\%. The measured PPS thickness is 1.35$\pm 0.05\, \mu$m, which is thin 
enough to use as a hydrogen marker. We have prepared 7 samples, S-1 to S-7, by stacking 
these Al foils and PPS films. Their details are  
listed in Table \ref{table-100}, together with the experimental data. 
It is notable that the total thicknesses of Al in the samples are more than 50\,$\mu$m 
to prevent He from entering the detector, taken into account the 43\,$\mu$m range 
of 8 MeV He in Al.  

The experimental setup and procedure are similar to the previous case \cite{Yam2019b}.  
A particle detector for energy analysis of H was set on the ion-beam axis and 
a sample was placed in front of the detector. The acceptance angular range of 
the detector, being expressed as  0 to $\phi_{\rm m}$ polar angle 
with respect to the beam axis, was adjusted by attaching a circular aperture 
on the detector. In the present experiments, we have chosen $\phi_{\rm m}=4.1^{\circ}$ 
for narrow and 11.1$^{\circ}$ for wide typical acceptance conditions 
using the apertures of 2.0 and 5.5 mm diameter, respectively. These are typical 
conditions chosen to obtain sufficient count rate of recoil H, depending on the 
H concentration in the sample. 

The recoil cross section for hard collision of $^{4}$He with $^{1}$H has a maximum at 
the incident He energy of 8--10 MeV \cite{Yam2019a} \cite{iba}. In this energy range, 
the stopping powers of matter with low atomic number such as Al decrease with 
increasing the He energy. This is also the case for recoil H in the MeV energy range. 
The depth resolution relying on the ion's energy loss 
degrades with decreasing stopping powers, namely with increasing the incident He energy 
in the 8--10 MeV range.  Therefore, we have chosen 8 MeV $^{4}$He$^{2+}$ energy to 
obtain relatively high count rate and high depth resolutions.
In the measurements, the beam current was $\sim$100\,pA for a beam spot of 
$\sim 50 \times 50\,\mu$m$^{2}$, which allowed 5--10 min. for accumulating a spectrum. 
The energy-measuring system for H was calibrated with 5.48 MeV $\alpha$ particles 
emitted from $^{241}$Am, together with a test pulser. 
The energy resolution of the measuring system including the particle detector was 32 keV.  

\section{ \large Results and discussion }\label{discuss}

\begin{figure}[h] 
\begin{center}
\includegraphics[width=120mm, bb=85 488 520 711]{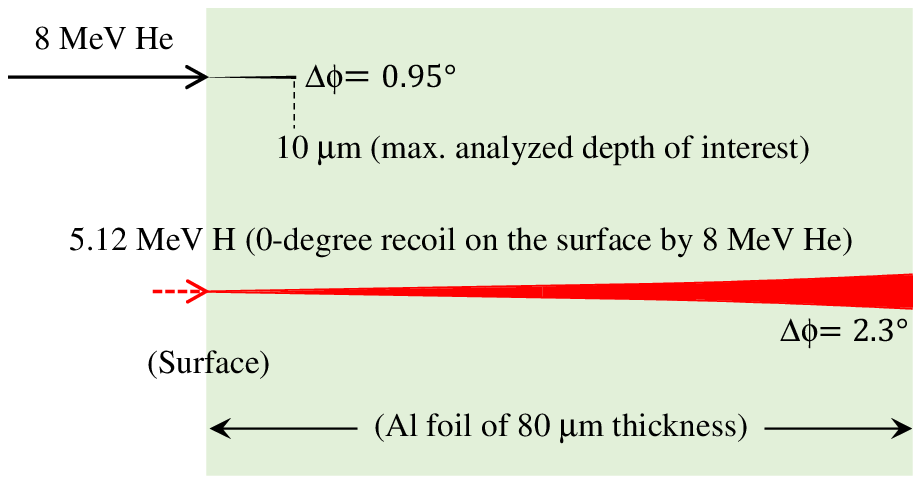}
\end{center}
\caption{Angular spreads $\Delta \phi$  calculated with TRIM for 8 MeV He 
and 5.12 MeV H (shown in red) incident on the Al foil of  80 $\mu$m thickness. 
The latter results from  head-on collision of 8 MeV He with H on the surface. 
For the Al foil of 50 $\mu$m thickness, $\Delta \phi = 1.2 ^{\circ}$ on the 
back surface. 
}
\label{fig-1}
\end{figure}

\hspace{5mm} It is of fundamental importance to confirm the applicability of a straight-path picture 
for the He and H trajectories in the present experimental conditions of T-ERDA. 
Figure \ref{fig-1} shows angular spreads of incident 8 MeV He and recoil H in the Al sample of 
80\,$\mu$m thickness, which were obtained by TRIM simulations \cite{Zie2013}. 
The angular spread of He in Al is as small as 0.95$^{\circ}$ from the beam axis at the depth of 10\,$\mu$m, 
which is the maximum analyzing depth of interest for typical use of the present T-ERDA.  
This depth is sufficiently shallower than the detection limit of 30\,$\mu$m depth, 
corresponding to zero energy of H after passing through the Al sample.   
The above angular spread causes uncertainty in the lateral direction of  
$10 \times 2 \tan 0.95^{\circ}=0.33\,\mu$m,  which is much smaller than the typical beam size 
of $\sim$3\,$\mu$m used for 3D mapping. Therefore, the lateral resolution at least up to the 
10\,$\mu$m depth remains approximately equal to the beam size.  
Meanwhile, most of recoil H starting from the Al surface in the 0$^{\circ}$ direction spread over 
a narrow angular range of  0--2.3$^{\circ}$ from the beam axis after passing through the Al foil. 

The straight-path picture confirmed above is applied to the motions of the projectile He, 
which is incident perpendicularly on the Al surface, and recoil H  
with a recoil angle $\phi \,  (0 \le \phi < \pi/2) $ from the beam axis.  In this case, 
the energy $E$ of recoil H measured after passing through the Al sample 
can be converted to the depth $z$ where H recoil occurs, i.e.,  
\begin{eqnarray}
E=\alpha \cos^{2}\phi \, 
\left( {\cal E_{\rm inc}}-\int_{0}^{z} {S_{\rm He}}\, dz \right) - \int_{z}^{L/\cos{\phi}} S_{\rm H}\, dz\:,
\label{eq-1}
\end{eqnarray}
where $\alpha=4 M_{\rm H}  M_{\rm He}/(M_{\rm H}+ M_{\rm He})^{2}=0.64$ with 
$M_{\rm H}$ and $M_{\rm He}$ being the atomic masses of H and He, respectively,
${\cal E_{\rm inc}}=8$\,MeV is the energy of incident He in the present case, 
$L$ is the thickness of the Al sample, $S_{\rm He}$ and $S_{\rm H}$ are 
the stopping powers of Al for He and H, respectively. The first term on the right side of 
Eq.\,(\ref{eq-1}) represents the energy of recoil H produced at $z$, while the second 
represents the energy loss of H in the outgoing path. 
In the following analysis, the values of stopping powers were obtained from the SRIM tables \cite{Zie2013}. 

\begin{figure}[h] 
\begin{center}
\includegraphics[width=120mm, bb=74 363 531 696]{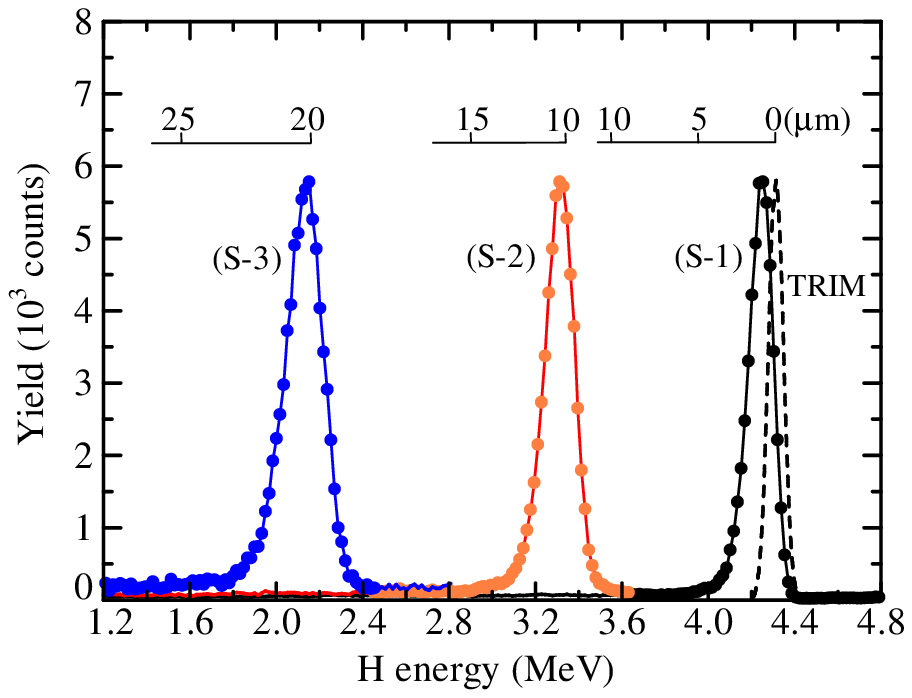}
\end{center}
\caption{T-ERDA spectra of H measured for S-1 to S-3 samples of Al/PPS/Al with 
the narrow acceptance of $\phi_{\rm m}=4.1^{\circ}$. The spectra have been scaled to 
have the same peak height. The depth scales based on the straight-path model, 
given by Eq.\,(\ref{eq-1}) for $\phi=0$, are indicated  for the three cases. 
The calculated spectrum using TRIM for 0-degree recoil is also shown with 
the same peak height for comparison of the peak width.  
} 
\label{fig-2}
\end{figure}

Figure \ref{fig-2} shows energy spectra of H measured under the narrow acceptance  
conditions ($\phi_{\rm m}=4.1^{\circ}$) for S-1 to S-3 samples of 
Al/PPS/Al(50 $\mu$m thickness). The thicknesses of the top Al layers are 
0, 10, and 20 $\mu$m for S-1, S-2, and S-3, respectively. 
For convenience of analysis, the yields have been 
scaled to have the same peak height as of S-1. We see sharp peaks originating from 
hydrogen in the PPS film. The three peaks are slightly asymmetric, which can be 
recognized by the larger number of plots (equally spaced with respect to H energy) 
on the left side than on the right side, indicating the existence of low-energy tails.  
The peak energies can be accounted for with the depth scales indicated in Fig. \ref{fig-2}, 
which were drawn with Eq.\,(\ref{eq-1}) for $\phi=0$. For the three cases, 
however, we see slight shift of the peak energy to the low-energy side from the scale.  
Such peak shift arises from the finite thickness of the PPS film. 
Since for both the incident He and recoil H, the ratio of PPS to Al stopping power is 0.64, 
the effective thickness of PPS measured on the Al depth scale is 
$1.35 \times 0.64= 0.86\,\mu$m. Accordingly, we may expect the peak shift equal to 
the half thickness of 0.43\,$\mu$m and thereby the observed peak shifts can be explained.
In Fig. \ref{fig-2}, the dashed curve shows the calculated spectrum by TRIM for 
0-degree recoil of H starting from the surface of 50-$\mu$m thick Al. 
The peak energy of 4.31 MeV by TRIM coincides with the origin of the depth scale 
indicated by using  Eq. (\ref{eq-1})  for $\phi=0$. About 60\% of the 
experimental FWHM value of 0.130 MeV corresponds to the TRIM value of 0.081 MeV 
which is due to energy straggling. 

\begin{figure}[h] 
\begin{center}
\includegraphics[width=120mm, bb=99 369 521 696]{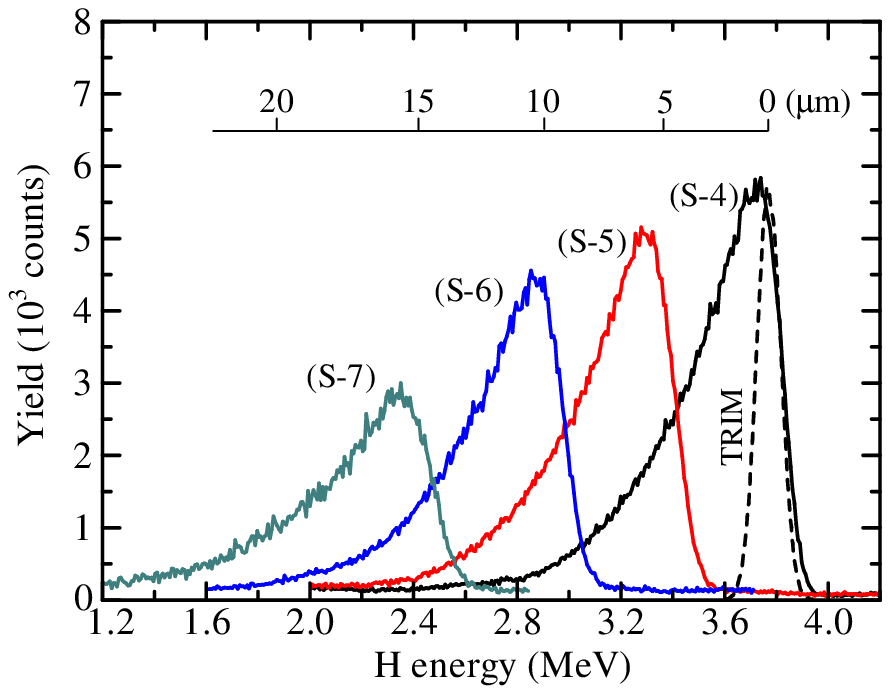}
\end{center}
\caption{T-ERDA spectra of H measured for S-4 to S-7 samples of Al/PPS/Al 
with the wide acceptance of $\phi_{\rm m}=11.1^{\circ}$.
The spectra are shown for the same number of incident He. 
The depth scale for Al based on the straight-path model, given by Eq.\,(\ref{eq-1}) 
for $\phi=0$, is indicated. 
The calculated spectrum using TRIM for 0-degree recoil (for the case 
shown in red in Fig. \ref{fig-1}) is also shown with the same peak height as of S-4
for comparison of the peak width. 
}
\label{fig-3}
\end{figure}

Figure \ref{fig-3} shows energy spectra of H measured under wide acceptance conditions 
($\phi_{\rm m}=11.1^{\circ}$) for S-4 to S-7 samples of Al/PPS/Al, 
in which H yields are shown for the same number of incident He. 
Low-energy tails are more discernible than those in Fig. \ref{fig-2}.  
We also see low-energy shift of the H peak from the depth scale, 
which is due to the PPS thickness, as already discussed for the case of narrow acceptance.   
In Fig. \ref{fig-3}, the dashed curve shows the calculated spectrum by TRIM for 
0-degree recoil of H starting from the surface of 80-$\mu$m thick Al 
(for the case shown in red in Fig. \ref{fig-1}). Again, the peak energy of  
3.77 MeV by TRIM coincides with the origin of the depth scale given by using  Eq. (\ref{eq-1})   
for $\phi=0$. The difference between the experimental FWHM value of 0.400 MeV for S-4 
and the TRIM value of 0.109 MeV is of essential interest in the following analysis. 
\begin{table}
\caption {Names and structures of the samples,  peak energies of H from the PPS layer in Al,  
FWHM values of the H peak, and  the depth resolutions $\Delta z$ determined after 
correction of the PPS thickness (1.35\,$\mu$m) to the FWHM values.  S-1 to S-3 are the
test samples for the narrow acceptance of the  detector  ($\phi_{\rm m}=4.1^{\circ}$), 
while S-4 to S-7 for the wide acceptance ($\phi_{\rm m}=11.1^{\circ}$). 
The uncertainty in the values of $\Delta z$ is $\pm$2.5\%.   }
 \vskip1.5mm
\begin{center}
 \begin{tabular}{lccccr}
 \hline\hline 
  & & \\[-5pt]
 Sample name and structure & Peak energy & FWHM &  FWHM & $\Delta z$ \\
   & (MeV) & (MeV) &  ($\mu {\rm m}$) & ($\mu {\rm m}$) 
\\[1.5mm] \hline 
   & & \\[-5pt]
  S-1: PPS$/$Al(50\,$\mu {\rm m}$) & 4.25 & 0.130 & 1.81 & 1.59 \\[1.5mm]
  S-2: Al(10\,$\mu {\rm m}$)$/$PPS$/$Al(50\,$\mu {\rm m})$ & 3.31 & 0.160 & 1.78 & 1.56 \\[1.5mm]
  S-3: Al(20\,$\mu {\rm m}$)$/$PPS$/$Al(50\,$\mu {\rm m})$ & 2.14 & 0.202 & 1.68 & 1.45 \\[2.5mm]
  S-4: PPS$/$Al(80\,$\mu {\rm {\rm m}}$) & 3.72 & 0.400 & 4.94 & 4.86 \\[1.5mm]
  S-5: Al(5\,$\mu {\rm m}$)$/$PPS$/$Al(75\,$\mu {\rm m}$) & 3.30 & 0.408 & 4.70 & 4.62 \\[1.5mm]
  S-6: Al(10\,$\mu {\rm m}$)$/$PPS$/$Al(70\,$\mu {\rm m}$) & 2.86 & 0.433 & 4.47 & 4.39 \\[1.5mm]
  S-7: Al(15\,$\mu {\rm m}$)$/$PPS$/$Al(65\,$\mu {\rm m}$) & 2.34 & 0.480 & 4.40 & 4.32 \\[1.5mm]
  \hline \hline
 \end{tabular}
\end{center}
\label{table-100}
\end{table}
 
The measured  FWHM values of H peaks for samples S-1 to S-7 are summarized in 
Table \ref{table-100}. The FWHM values in MeV have been converted to those in $\mu$m, by 
applying the depth scales indicated in Figs. \ref{fig-2} and \ref{fig-3}. In Table \ref{table-100}, 
the depth resolutions $\Delta z$ have been estimated by subtracting the effect of PPS 
thickness by assuming the convolution characteristic of Gaussian distributions, i.e., 
$\Delta z=(1.81^{2}-0.86^{2})^{1/2}=1.59\,\mu$m for S-1, for example. 
For S-4, the calculated value of mean energy spread 0.109 MeV due to 
energy straggling, noted earlier, corresponds to the depth uncertainty of 1.33$\,\mu$m, 
which might be the best depth resolution that can be reached in this case. 

For S-1 to S-3 in Table \ref{table-100}, FWHM (MeV) increases with 
increasing the thickness of Al foil on PPS, whereas FWHM ($\mu$m) and $\Delta z$ decrease 
oppositely.   Similar opposite dependences are also seen for S-4 to S-7.  
Such behavior is expected to occur for two reasons. First, the stopping powers of Al for both 
the incident He and recoil H increase as they lose energy in Al, which  
extends the depth scale with increasing depth. Second, the peak widths are predominantly 
determined by a different factor from the energy-loss process, which will be discussed later, 
otherwise the peak width should increase proportionally to the depth scale. 
Such feature of T-ERDA simplifies understanding of 3D images of H distributions because 
the depth resolution does not degrade with increasing depth, but it rather remains  
approximately constant in the depth range of interest, as is seen in Table \ref{table-100}.  

\begin{figure}[h] 
\begin{center}
\includegraphics[width=120mm, bb=67 388 505 701]{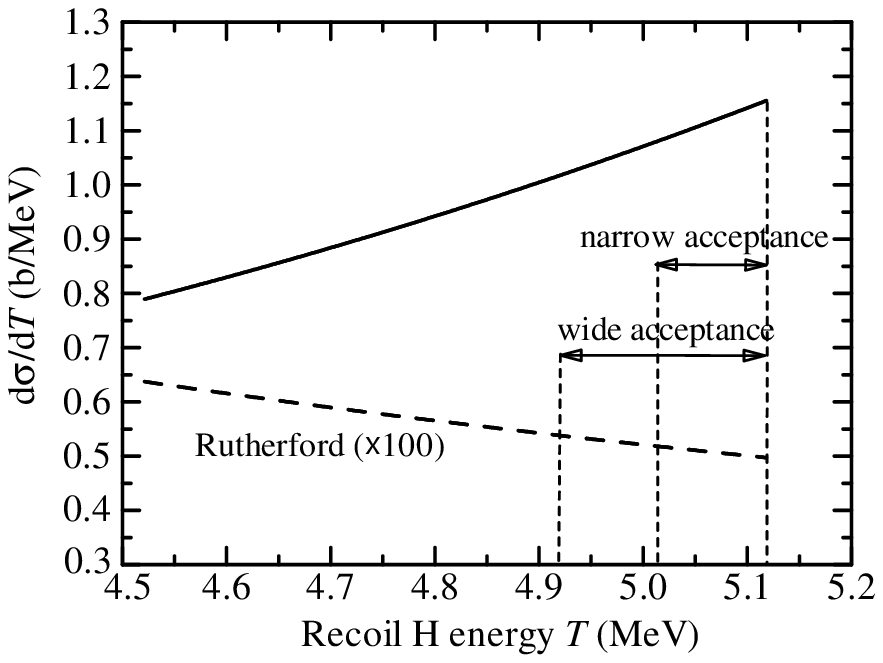}
\end{center}
\caption{Recoil cross section as a function of recoil H energy, representing 
the initial energy spectrum of recoil H.  The dashed curve shows the Rutherford case 
(Coulomb interaction only) for comparison. The energy ranges corresponding to 
the narrow ($\phi_{\rm m}=4.1^{\circ}$) and wide  ($\phi_{\rm m}=11.1^{\circ}$) 
acceptance conditions are indicated. 
}
\label{fig-4}
\end{figure}

\begin{figure}[h] 
\begin{center}
\includegraphics[width=120mm, bb=68 385 497 698]{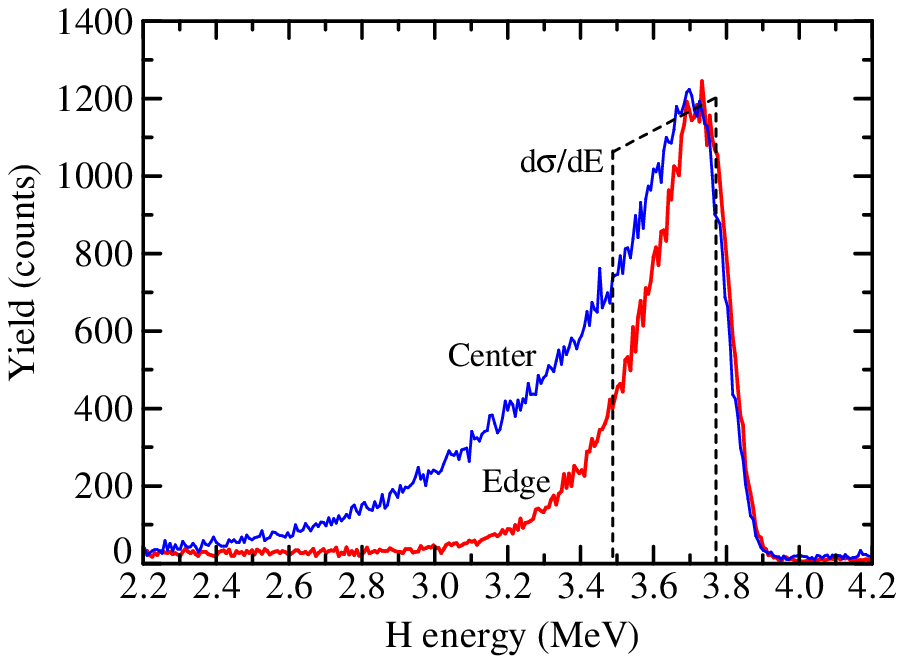}
\end{center}
\caption{T-ERDA spectra for incidence of 8 MeV He on the center 
and edge of S-4 sample. In the latter case, less than about one-side half of 
the circular window of the detector is in the shadow of the sample holder, 
thereby low-energy H is selectively prevented from entering the detector. 
The calculated cross section $d\sigma / dE$  is 
also shown to have the same peak height. 
}
\label{fig-5}
\end{figure}

To understand the origin of low-energy tails of H peaks appearing remarkably 
under the wide acceptance conditions, we need to follow the processes of recoil and 
energy loss.  Usually, the recoil cross section $d \sigma / d \Omega$ is written 
as a function of  $\phi$. By variable transformation from $\phi$ 
to the recoil H energy $T$ using the relation
\begin{eqnarray}
T=\alpha {\cal E_{\rm inc}}\cos^{2} \phi \;,
\label{eq-2}
\end{eqnarray}
the recoil cross section $d \sigma /dT$, which represents the  initial energy 
distribution of recoil H, is expressed as
\begin{eqnarray}
\frac{d\sigma}{d T}=\frac{d\sigma}{d \Omega}\,\frac{d \Omega}{d T}
=\frac{d\sigma}{d \Omega}\,\frac{\pi}{\sqrt{\alpha {\cal E_{\rm inc}} T}}\;,
\label{eq-3}
\end{eqnarray}
where $d \Omega =2\pi\sin \phi\, d \phi$.  
Figure \ref{fig-4} shows $d \sigma / dT$ calculated from Eq. (\ref{eq-3})
using $d\sigma / d \Omega$ given by SigmaCalc data \cite{SigCal}, together with 
that for Coulomb interaction only, i.e., for the case of Rutherford scattering. 
The energy ranges which correspond to the angular acceptance range of the 
detector, under the straight-path assumption, are shown by arrows. 
From Eq. (\ref{eq-3}), the observed H-energy spectrum is given by 
\begin{eqnarray}
\frac{d\sigma}{d E}=\frac{d\sigma}{d T}\,\frac{d T}{d E}\;. 
\label{eq-4}
\end{eqnarray}
For Al of 80 $\mu$m thickness, $dT/dE=0.68$ in the ranges of $T$ and $E$ 
in consideration is obtained numerically using Eq. (\ref{eq-1}). 
The calculated $d\sigma /dE$ for the case of S-4 
($z=0, \phi_{\rm m}=11.1^{\circ}$, and $L=80\,\mu$m) is shown in Fig. \ref{fig-5}.  

High- and low-energy sides of the H peak originate from the energy signals 
generated from the central and peripheral area of the detector window 
(5.5 mm diameter for S-4, as noted in \S\ref{exp}), 
respectively. This is evident from the two spectra for S-4, shown in Fig. \ref{fig-5}, 
which were measured for the same number of incident He ($1.4 \times10^{11}$). 
As the beam spot was moved from the center to the edge of the sample mounted on 
the sample holder, the tail on the low-energy side reduced drastically, as seen in 
Fig. \ref{fig-5}. This is because the peripheral area of the detector window is 
partly in the shadow of the sample holder, preventing low-energy H in the 
shadow from entering the detector. It should be noted that the tail of the spectrum 
for the center, shown in blue in Fig. \ref{fig-5}, is slightly higher than the corresponding 
spectrum for S-4 in Fig. \ref {fig-3}, although the FWMH values obtained from these 
different runs are approximately the same. This demonstrates sensitive dependence 
of the tail on the experimental conditions, for example, a possible deviation 
of the center of the detector from the beam axis. Considering the above 
correspondence between the H energy and the position on the detector window, 
the low-energy side of the H peak should be smeared considerably by the angular 
spread of recoil H during passage in Al, in contrast to the high-energy side. 
This causes the low-energy tail, giving rise to the asymmetric triangular spectrum. 
  
In the following, the representative cases for S-1 and S-4 are discussed since 
general aspects of the peak width can be deduced from these two cases. 
According to calculations with known stopping powers, the 
range of $5.094 \le T\, {\rm (MeV)}\le5.120$ for the narrow acceptance leads to 
$4.277 \le E\, {\rm (MeV)}\le 4.309$ after passing through Al of 50\,$\mu$m thickness. 
Similarly, $4.93 \le T\, {\rm (MeV)}\le 5.12$ for the wide acceptance 
leads to $3.49 \le E\, {\rm (MeV)}\le 3.77$ after passing through Al of 
80\,$\mu$m thickness. The widths of these $E$ ranges, 0.032 and 0.28\, MeV,  
can be increased and smeared by two intrinsic factors. One is the energy 
straggling of H in Al, which is recognized as the FWHM peak widths of the TRIM spectra, 
0.081 and 0.109 MeV in Figs. \ref{fig-2} and \ref{fig-3}, respectively. 
Another factor is the counting of low-energy recoil H with a recoil angle  
greater than $\phi_{\rm m}$, that can enter the detector after being deflected 
towards the detector. Actually, the FWHM angular spreads of H after passing 
through Al of 50 and 80 $\mu$m thicknesses are 1.2 and 2.3$^{\circ}$ from 
the beam axis, respectively.  This means that the low-energy ends of the $T$ ranges 
noted above should be replaced by the convoluted values of 
$(4.1^{2}+1.2^{2})^{1/2}=4.3^{\circ}$, and by $(11.1^{2}+2.3^{2})^{1/2}=11.3^{\circ}$ 
for the narrow and wide acceptance, respectively. These values lower the low-energy ends 
of the $T$ ranges noted above, providing the modified widths of $E$ ranges, 0.035 and 
0.32\, MeV, instead of 0.032 and 0.28\, MeV, for the narrow and wide acceptance, 
respectively. We again apply the convolution characteristic 
to the four energy widths, i.e., the energy acceptance, the width due to energy straggling, 
the energy resolution of the detector system 0.032 MeV, and the energy loss of H during 
passage through the PPS layer 0.013 MeV, although the last one makes little contribution 
to the results.  Finally, we obtain the energy widths of 
$\Delta E= (0.035^{2}+0.081^{2}+0.032^{2}+0.013^{2})^{1/2}=0.095$ and 
$(0.32^{2}+0.109^{2}+0.032^{2}+0.013^{2})^{1/2}=0.34$ MeV 
for the narrow and wide acceptance, respectively. 
These values are 73 and 85\% of the experimental values of 0.130 and 0.400 MeV 
for S-1 and S-4 samples, respectively. Such fair agreement between the experimental 
and calculated values is rather satisfactory, considering the crudeness of 
the present analysis. From the above discussion, the following conclusion is obtained. 
For narrow acceptance conditions, the main factor to degrade the energy resolution, 
and therefore, the depth resolution is the energy straggling of H in Al. As the angular 
acceptance of the detector increases, the low-energy side of the spectrum extends 
to form a tail, while the high-energy side remains unchanged. This causes the degraded 
depth resolution corresponding to the deformed triangular spectrum.  
  
In relation to the spectrum shape, it is notable to compare the present nuclear elastic 
condition with the Rutherford case, shown in Fig. \ref{fig-4}.  
In addition to the extremely larger cross section than the Rutherford case, there is a 
noticeable difference between the two. Indeed, $d \sigma / dT$ in the present case decreases with 
decreasing $T$, while it increases as $d \sigma / dT \propto T^{-2}$ in the Rutherford case.  
Accordingly, the high-energy recoil more contributes to the high energy side of 
the H peak in the nuclear elastic case than in the Rutherford case. 
This certainly causes sharp high-energy side of the H peak in the nuclear elastic condition. 

In the previous work under wide acceptance conditions of 15.3$^{\circ}$, we reported 
that the depth resolution of $\sim$1.2 $\mu$m was estimated using the high-energy edge of 
the H image \cite{Yam2019b}, without noticing the long tail that must be hidden under 
the H image. Therefore, the above value must be corrected to a larger value 
of more than 4.86 $\mu$m for S-4 under the present definition of the depth resolution. 
However, the spectrum shape having a long tail cannot be 
fully characterized only by its FWHM value. The contrast of 3D image, like those reported 
previously, might be more suitably understood with the sharpness of high-energy edge of the 
image rather than with the FWHM value. Thus, the depth resolution in T-ERDA should be 
properly defined, depending on the observed image of interest. 

\section{ \large Conclusion }\label{conclude}

\hspace{5mm} The feature of the depth resolution in T-ERDA has been studied for a better 
understanding of 3D images of H distribution in matter. The main degrading factor for 
the depth resolution is the energy straggling of H in Al for narrow acceptance conditions 
of the detector. As the acceptance angle of the detector increases, the low-energy 
side of recoil H spectrum extends to form a tail, while the sharp high-energy side remains 
unchanged. It follows that the depth resolution is degraded by the enhanced FWHM 
value of the deformed triangular spectrum.  
 
In T-ERDA using a micro-beam, 3D images of H distributions in a local structure in the 
sample can be obtained by energy-to-depth conversion using the local stopping power 
associated  with the local atomic composition.  Therefore, such 3D images possibly inform 
us of the atom species existing around H in the sample. 
3D images of H-containing aggregates possibly allow, for example, to identify abundances 
and distributions of water or OH in minerals and natural glass samples, based on 
the knowledge about depth resolutions obtained in the present work. 
For such analysis, simultaneous measurements of characteristic x-rays induced by 
the He beam will be helpful \cite{Yam2019a}. \\
\newpage
\hspace{-7.5mm} {\bf ACKNOWLEDGEMENTS}\\
\indent The authors thank the technical staff of UTTAC for operating the 
6MV tandem accelerator, and thank G. Kawasaki and S. Hatada for their support 
during experiments. 
We also thank Dr. M. Kurosawa for valuable comments for the manuscript. 
We are grateful to Toray KP Films Inc. for providing the PPS films, and to 
Toyo Aluminium K.\,K. for providing the high-purity Al foils. 
This work was developed partly from the experimental research program supported by 
Cross-Ministerial Strategic Innovation Promotion Program -- Unit D66 --Innovative 
measurement and analysis for structural materials 
 (SIP-IMASM) operated by the Japanese cabinet office.


\newpage
\newpage
\newpage
\newpage
\newpage

\end{document}